\documentclass[]{aastex631}

\usepackage{graphicx}
\usepackage{dcolumn}
\usepackage{bm}
\usepackage{amsmath}	
\usepackage{tikz}
\usepackage{cancel}
\usepackage{braket}

\newtheorem{theorem}{Theorem}[section]

\newtheorem{proposition}{Proposition}[section]
\newtheorem{remark}{Remark}[section]
\DeclareMathAlphabet{\pazocal}{OMS}{zplm}{m}{n}

\submitjournal{ApJ}

\begin{document}


\title{What if the universe expands linearly? \\ A local general relativity to solve the ``zero active mass'' problem}

\author[0000-0003-3100-2394]{Robert Monjo}
\affiliation{Department of Algebra, Geometry and Topology, Complutense University of Madrid\\ Pza. Ciencias 3, E-28040 Madrid, Spain, rmonjo@ucm.es}



\begin{abstract}
Modern cosmology presents important challenges such as the \textit{Hubble tension}, \textit{El Gordo's collision} or the \textit{impossible galaxies} ($z > 10$). Slight modifications to the standard model propose new parameters (e.g. the early and dynamical dark energy). On the other hand, alternatives such as the coasting universes (e.g. the \textit{hyperconical model} and the spatially flat $R_h = ct$ universe) are statistically compatible with most of observational tests, but still present theoretical problems in matching the observed matter contents since they predict a ``zero active gravitational mass.'' To solve these open issues, we suggest that general relativity might be not valid at cosmic scales, but it would be valid at local scales. This proposal is addressed from two main features of the embedding hyperconical model: 1) the background metric would be independent of the matter content, and 2) the observed cosmic acceleration would be fictitious and because of a distorted stereographic projection of coordinates that produce an apparent radial inhomogeneity from homogeneous manifolds. Finally, to support the discussion, standard observational tests were updated here, showing that the hyperconical model is adequately fitted to Type Ia Supernovae, quasars, galaxy clusters, BAO, and cosmic chronometer data sets. 
\end{abstract}



\section{Introduction}


\subsection{Motivation}

General relativity (GR) is commonly assumed to be valid at cosmic scales, but it might not be. After the breaking observations of ``Hubble tension'' \citep{Poulin18, Poulin19, KamionkowskiRiess}, ``El Gordo's collision'' \citep{Zhang_2015, Asencio_2020, Asencio_2023} and the ``impossible early galaxies'' ($z > 10$) \citep{Yennapureddy2018, Ferreira2022, Gupta2023, McQuinn_2024}, deep discussions on the standard $\Lambda$CDM model and modified cosmology theories have intensified. Thus, the need to explore alternative models should not be discarded, even if it is limited to a minority sector \citep{kroupa_2013, Merritt2017}. An alternative is the linear expansion described by the nonempty \textit{coasting universes} \citep{Melia2018, John2019, MCS2023}, first proposed by R. Kolb \citep{Kolb1989, John1997, John2000} and deeply studied by Melia under the name $R_h=ct$ for its flat-space case \citep{Melia2007, Melia2016, Melia2016b, Melia2017, Melia2018, John2019}. The main advantage of the simple linear expansion hypothesis is that it can explain some conflicts or tensions observed with respect to the standard model \citep{Melia2018, Melia2019, Melia2022, Melia_Lopez2022}. Another recent finding is that linear expansion and $\Lambda$CDM models fit equally well to the gravitational-wave standard sirens observed in the LIGO-Virgo-KAGRA detector \citep{Raffai_2023}. 

Despite these successful results, the linear expansion faces some challenges. For instance, $R_h = ct$ presents difficulties in explaining the observed abundance of helium about $\mathrm{He/H} \sim 0.244$ in mass, presumably produced by the ``Big Bang nucleosynthesis'' or primordial nucleosynthesis in the early universe \citep{Lewis2016, Chardin2021}. An alternative explanation for that helium mass fraction is hydrogen fusion in primordial stars, which predicts a black body temperature of about $T = 2.76 K$, very close to the observed temperature for the Cosmic Microwave background (CMB) radiation \citep{LopezCorredoira2017}. Another flagship of the standard model, the anisotropic temperature distribution in the CMB, could be produced at $z \sim 10$ instead of the standard $z \sim 1080$ due to dust created in Population III stars \citep{Vavrycuk2017, Melia2022b}. 

However, some points seem to still be favored by the $\Lambda$CDM model, such as the matter content \citep{Mitra2014}, the Type Ia supernova luminosity distance-redshift relation \citep{Bilicki2012, Shafer2015, Hu2018, LopezC2022}, the first BAO peak position \citep{Tutusaus2016, Planck2020a, Planck2020b, Fujii2020}, and high-redshift quasars \citep{Bilicki2012, Fan2023}, among others. To solve these tensions, \citet{Monjo2018} proposed a geometric link between linear and accelerated universes that emerges from the hyperconical model, developed in \citep{Monjo2017, MCS2020, MCS2023}.


\subsection{Challenges of modern cosmology}

The standard model requires a remarkable number of parameters to solve well-known classical issues such as: (1) the ``horizon problem'' on the homogeneity of regions that were (hypothetically) causally disconnected from themselves, (2) the ``flatness problem'' regarding why the total density is ``exactly'' equal to the critical density that separates open and closed manifolds, and (3) the ``synchronicity problem'' on the unity $H_0t \sim 1$ given by the (almost) coincidence between the Hubble time ($1/H_0$) and age $t$ of the universe just now \citep{Avelino2016, MCS2023}. \textit{Inflation theory} attempts to solve the first two problems (horizon and flatness) by using a hypothetical exponential expansion in the early stages of the universe ($< 10^{-32}$ seconds) because of a kind of \textit{dark energy}, which could favor the high homogeneity and isotropy of the universe today \citep{Guth1981}. After the appearance of the Hubble tension, a new parameter entered the playing field: the early dark energy (EDE), which could have a role in the early stages of the universe \citep{KamionkowskiRiess, Herold2023}. However, new physics could also be necessary for the late-time universe \citep{Hu2023}.

Dynamical dark energy is a generalized idea of different values of late--early dark energy \citep{Peebles2003}, but with more general dynamics such as $\Omega_\Lambda \propto a^{-3(1+w(z))}$ for some variable equation of state $w(z) \neq -1$ \citep{Mainini_2003, Zhao2017, Zhang2023, DiValentino2023}. For instance, \citep{Zhao2017} used a large set of observations to constrain the equation of state $w(z)$ of dark energy and found that it oscillates between $-0.4$ and $-2$, approximately, for $z \in (0,2)$. \citet{Zhang2023} proposed a model with a small deviation with respect to the standard cosmology during the past and a faster expansion in the future (i.e., at $z < 0$). On the other hand, \citet{Mainini_2003} separately modeled slow (classical) or rapid (supergravity) growth, leading to $w(z) \sim (-0.3, -0.6)$ and $w(z) \sim (-0.4, -0.9)$, respectively, both evolving from $z = 10$ to $z=0$. Meanwhile, \citep{DiValentino2023} modeled $w(z) \in (-0.4, -1.5)$ for $z \in (0, 0.5)$ and steadily shifted around $w(z) \sim -1$ for $z > 1$. 

Another modification of the standard model is assuming that dark energy and dark matter can non-minimally couple to spacetime (scalar) curvature \citep{Hussain2023}, which produces stable dynamics from the new cosmological models. However, the dark-matter concentration is only consistent with the Planck mission \citep{Planck2020a} when self-interacting dark matter is added to adequately fit the galaxy rotation curves \citep{Ren2019, Girmohanta2023}. The advantage of the \textit{dark matter halo} models is that they use fitting parameters to represent the diversity of galaxy dynamics better than the only Milgromian constant used in modified Newtonian dynamics (MOND) \citep{Ren2019, Milgrom2020, Rodrigues2023}. However, dark matter is not directly detected yet \citep{Abel2017, Du2022}, and models based on this paradigm have serious difficulties explaining galaxy cluster collisions like ``El Gordo'' \citep{Asencio_2020, Asencio_2023}, or the formation of (early) large structures such as the Giant Arc \citep{Lopez2022} and the Big Ring \citep{Lopez2024}, among others \citep{Haslbauer2020, Aluri2023}.

\subsection{Beyond the dark energy and dark matter}

Several authors proposed alternative models based on linear expansion. For instance, the Dirac--Milne or Milne model was proposed in 1933 as a solution for the Einstein field equations $G_{\mu\nu} = 8\pi\mathrm{G}T_{\mu\nu}$ for empty universe $T_{\mu\nu}=0$ with the Friedmann--Lamaître--Robertson--Walker (FLRW) metric \citep{Gehlaut2003, Levy2012}. This model leads to a linear expansion (i.e. scale factor $a \propto t$) with negative spatial curvature (i.e., $k = -1$), so it does not need the inflation stage. Alternatively, coasting models (with $k = 0$, $+1$ or $-1$) define nonempty universes by a ``zero active (gravitational) mass'' \citep{John2000, Melia2017, MCS2023}, an exotic $K$-matter \citep{Kolb1989, Gehlaut2002}, or a dynamical dark energy (varying as $a^{-2}$ like the curvature term), but with a constant equation of state of $w(z) = - \frac{1}{3}$ to obtain a linear expansion during the whole cosmological time or at least during its predominance stages \citep{John1997, MCS2020, MCS2023}. The ``zero active gravitational mass'' was sometimes interpreted as a result of a matter-antimatter symmetric universe \citep{Villata_2011, Hajdukovic2019, Chardin2021} but is incompatible with observations because they should annihilate themselves and, moreover, they do not produce the required (repulsive) antigravity \citep{Cohen_1998, Anderson2023}. Despite \citet{Melia2018} showing a presumed (observational) superiority of $R_h=ct$ over the standard model, some doubts remain about the physical interpretation of its ``apparently empty'' or ``zero active gravitational mass'' \citep{Mitra2014}.

This paper aims to reinterpret the $R_h=ct$ universe with its geometrical link to the standard model according to hyperconical perspectives. To establish the notation, the structure of the paper begins with Section \ref{sec:standard_and_coasting}, which briefly reviews the standard (accelerated) and coasting (inertial) universes. Section \ref{sec:hyperconical} summarizes the development of the hyperconical model, including the statement of the fictitious acceleration. Section \ref{sec:observational} collects some observational tests by using data from Type Ia Supernovae (SNe), cosmic chronometers (CC), baryon acoustic oscillations (BAO), quasars, and galaxy clusters. Section \ref{sec:refoundation} suggests the need for a re-foundation of cosmology and provides an interpretation of the ``zero active mass'' within the new paradigm. Finally, Section \ref{sec:perspectives} sums how hyperconical perspectives lead to both $R_h=ct$ and standard-like metrics. 

To support our main statements, three appendices follow the paper as supplementary material: {Appendix}~\ref{sec:embedding} develops the derivation of the hyperconical metric; {Appendix}~\ref{sec:accel} proofs that its apparent radial inhomogeneity can be assimilated to a fictitious acceleration compatible with the standard model; and {Appendix}~\ref{sec:dataCC} is reserved to share the data used in Section~\ref{sec:cc}.

\section{Accelerated and linear expansions}
\label{sec:standard_and_coasting}

\subsection{Homogeneous accelerated universes}
\label{sec:standard}

Standard-based K$\Lambda$CDM models are characterized by the cosmological principle on homogeneity and isotropy at large scales, thus they are represented by the Friedmann--Lamaïtre--Robertson--Walker (FLRW) metric, $g_{{}_{FLRW}} = (g_{{}_{FLRW}})_{\mu\nu} dx^\mu dx^\nu \equiv ds^2_{FLRW}$,  usually expressed in coordinates as follows: \begin{eqnarray} 
ds^2_{FLRW} = dt^2
-  a(t)^2\left( \frac{dr'^2}{1-Kr'^2} + {r'}^2d{\Sigma}^2 \right)\,.
\end{eqnarray}where $r'$ is the comoving distance, $t$ is the coordinate ``time,''  $\Sigma$ represents the angular coordinates, $K \equiv K_{FLRW}$ is the spatial curvature, and $a(t)$ is the scale factor. Then, the Ricci curvature scalar is\begin{eqnarray} \label{eq:R_stand}
    R_{FLRW} = - \frac{6}{a^2}(\ddot{a} a + \dot{a}^2+ K)
\end{eqnarray} and Friedmann equations are finally obtained from Einstein field's equations by assuming a perfect homogeneous fluid at rest in the FLRW metric: 
\begin{eqnarray} \label{eq:Friedmann.eq1}   
\left(\frac{\dot{a}}{a} \right)^2 &=& \frac{8{\pi}{\mathrm{G}}\rho}{3} - \frac{K}{a^2}   
\\
 \label{eq:Friedmann.eq2}   
\left(\frac{\ddot{a}}{a} \right)  &=& - \frac{4{\pi}{\mathrm{G}}}{3}(\rho+3p) 
\end{eqnarray}where $\rho = \rho_{\mathrm{m}}+\rho_{\mathrm{r}}+\rho_{\mathrm{\Lambda}}$ and $p = p_{\mathrm{m}}+p_{\mathrm{r}}+p_{\mathrm{\Lambda}}$ are total density and pressure, including matter (m), radiation (r), and dark energy ($\Lambda$) with $\rho_{\Lambda} = - p_{\Lambda} := \Lambda/(8\pi\mathrm{G})$. Notice that linear expansion ($a = t/t_0$) is found for the condition of $\rho+3p=0$, which is equivalent to equation of state $w := p/\rho = -\frac{1}{3}$ for $\rho \neq 0$ \citep{Melia2012}. However, if the universe is empty ($p = \rho =0$), the Milne--Dirac model is obtained with $K_{FLRW}= -1$ and $a = t/t_0$ \citep{Gehlaut2003, Levy2012}.

Defining the Hubble parameter as $H := \dot{a}/a$ and identifying the critical density defined by $\rho_{crit} := {3{H_{0}}^{2}}/{8\pi G}$, where $H_{0}$ is the current value for the Hubble parameter, Eq.~\ref{eq:Friedmann.eq1} is usually rewritten as follows:
\begin{eqnarray} \label{eq:Friedmann.eq1b}   
\frac{{H}^2}{{{H}_{0}}^2}  = \frac{\rho_r}{{\rho}_{crit}} + \frac{\rho_m}{{\rho}_{crit}} + \frac{{\rho}_{\Lambda}}{\rho_{crit}} - \frac{K}{{{H}_{0}}^2 a^2} 
\end{eqnarray} Assuming that energy density is dominated by (non ultra-relativistic) cold matter, $\rho \sim \rho_m$, its equation of state $p = w\rho$ is approximately $w \approx 0$ and it varies by the expansion as $a^{-3}$. If the universe is dominated by radiation or ultra-relativistic particles, $w \approx 1/3$ and $\rho \sim \rho_r$ varies as $a^{-4}$.  Therefore, Eq.~\ref{eq:Friedmann.eq1b} is rewritten as follows:
\begin{eqnarray} \label{eq:Friedmann.eq1c}   
\frac{{H}^2}{{{H}_{0}}^2}  = {\Omega}_{r} \left(\frac{{a}_{o}}{a} \right)^{4} + {\Omega}_{m} \left(\frac{{a}_{o}}{a} \right)^{3} + {\Omega}_{K} \left(\frac{{a}_{o}}{a} \right)^{2} + {\Omega}_{\Lambda} \;\;\; 
\end{eqnarray} where ${\Omega}_{i} := {{\rho}_{i}}/{{\rho}_{crit}}$ are the $\Lambda$CDM parameters for radiation (${\rho}_{r}$), matter (${\rho}_{m}$), dark energy (${\rho}_{\Lambda}$), and curvature (${\Omega}_{K} := -K/H_0^2{a}_{o}^2$). Because redshift $z$ depends on the scale factor as $1+z = {a}_{o}/a$, trivially, the parameter Hubble $H$ depends on the redshift $z$. International projects assuming the $\Lambda$CDM model, such as the Planck Mission, have estimated that ${\Omega}_{r} \approx 9 \cdot 10^{-5}$, ${\Omega}_{K} \approx 0$, ${\Omega}_{m} \approx 0.3$, and ${\Omega}_{\Lambda} \approx0.7$ (\citealp[e.g., ${\Omega}_{m} = 0.3111  \pm 0.0056$, ${\Omega}_{\Lambda} = 0.6889 \pm 0.0056$ according to Table 2 of ][]{Planck2016}). Finally, the deceleration parameter is defined as follows: \begin{eqnarray} \label{eq:desac}
    q = - \frac{\ddot a a}{{\dot a}^2} = -1 - \frac{\dot H}{H^2}\,,
\end{eqnarray}which is now $q_0 \approx \frac{1}{2}\Omega_m - \Omega_\Lambda \approx -1 +\frac{3}{2}\Omega_m \approx -0.54$ for ${\Omega}_{r} \approx 0 \approx {\Omega}_{K}$.  Then, the Ricci curvature scalar is\begin{eqnarray} \label{eq:R_stand2}
    R_{FLRW} = - \frac{6\dot{a}^2}{a^2}\left(\frac{\ddot{a} a}{\dot{a}^2} + 1\right) = - 6H^2\left(1 - q\right)
\end{eqnarray} 
Because of the interest in this work, the parameter of curvature ${\Omega}_{K}$ has not been neglected.

\subsection{Coasting universes and linear expansion}
\label{sec:coasting}

The first linearly expanding universe, or \textit{eternal coasting model}, was proposed by \citet{Kolb1989}, who suggested a predominant exotic $K$-matter, whose equation of state is $p_K = -\rho_K/3$, leading to a dynamical evolution proportional to $a^{-2}$ like a spatial curvature \citep{Gehlaut2002, Gehlaut2003}. However, this model is characterized by a large relativistic epoch with a predominance of radiation (varying as $a^{-4}$), and hence its expansion rate is not uniform. On the other hand, \citet{John1996, John1997} proposed a closed cosmological model that coasts throughout cosmic time after the Planck epoch, in contrast to the Ozer-Taha model, which reaches $a \propto t$ for some time, but soon transitions to a decelerating standard evolution \citep{OzerTaha1986}. As an alternative to the theory of inflation, these models solve both the flatness and the horizon problems by bouncing, that is, a nonsingular solution
with $a = \sqrt{a_0^2 + t^2}$ for some small radius $a_0$ of the universe. \citet{John1996, John1997} modeled this behavior for the entire cosmological history, with a minimum radius $a_0$ of the order of Planck length.  Except for the short Planck epoch, the coasting models provide an inertial or eternal linear expansion under the $FLRW$ metric, that is:\begin{eqnarray}
   ds^2_{coast} \approx dt^2 
-  \frac{t^2}{t_{0}^2} \left( \frac{dr'^2}{1-Kr'^2} + {r'}^2d{\Sigma}^2 \right) 
\end{eqnarray} and the $R_h = ct$ metric can be taken as a particular case \citep{Melia2007, Melia2012, John2019}, with $K_{FLRW} = 0$: \begin{eqnarray}
   ds^2_{R_h=ct} \approx dt^2 
-  \frac{t^2}{t_{0}^2} \left({dr'^2} + {r'}^2d{\Sigma}^2 \right) 
\end{eqnarray}Because $a = t/t_0$, both models have the same cosmological timeline, with a linear Hubble parameter $H = \dot{a}/a = 1/t = t_0^{-1}a = H_0\,(1+z)$. In units of $H_0 \equiv 1$, that is\begin{eqnarray}
    H_{coast}(z) \;\; = \;\; 1\; + \;z 
\end{eqnarray}

\smallskip
\smallskip
\smallskip

\section{Hyperconical universes}
\label{sec:hyperconical}

\subsection{Description of the model}

Linear expansion is naturally obtained by embedding our universe in an ambient manifold. Let $\eta$ be the 5-Minkowskian metric with signature $(1,4)$ and $\pazocal{H}^4 := \{ X \in \mathbb{R}^{1,4}_{\eta} \; :\;  \left|X - O \right|_{\eta} = \beta_0 t \}$ be a hypercone for some constant $\beta_0$, time coordinate $t$ and origin $O$ of the hypercone. If (comoving) observers on its hypersurface are considered, the dynamical embedding $\mathcal{T}_{t}$ of $\pazocal{H}^4 \subset \mathbb{R}^{1,4}_{\eta}$ leads to a metric $g$ that is similar to the FLRW metric except for the lapse and shift terms, which produce a radial inhomogeneity in the space \citep{Monjo2017, Monjo2018}. Locally, the differential line is (Appendix \ref{sec:embedding}): \begin{eqnarray}  \label{eq:inhom0}
ds^2_{hyp} \approx dt^2 \left(1- kr'^2 \right) 
-  \frac{t^2}{t_{0}^2} \left( \frac{dr'^2}{1-kr'^2} + {r'}^2d{\Sigma}^2 \right)\; - \; \frac{ 2r't}{t_{0}^2} \frac{dr'dt}{\sqrt{1-kr'^2}} \;\;\;\;
\end{eqnarray}where $k^{-1} := (1-\beta_0^2)t_0^2$ is the squared curvature radius link to a curvature $k \equiv k_{hyp} \neq K_{FLRW}$, while $a := t/{t_0}$ is a scale factor, $r' << t_0$ is the comoving distance, $\Sigma$ represents the angular coordinates, and $t_0 \equiv 1$ is the current value for the age $t$ of the universe. Both the Ricci scalar of curvature and Friedmann equations derived from this universe for $k = 1 = 1/t_0^2$ are locally equivalent to those obtained for a spatially flat $\Lambda$CDM model with linear expansion. Particularly, the local Ricci scalar is, at every point ($r'\equiv 0$) equal to \citep{Monjo2017}:\begin{eqnarray}
    R_{hyp} \approx -\frac{6}{t^2} =  R_{FLRW}\bigg|_{K\,=\, 0,\; a\,=\,t/t_0}
\end{eqnarray}as for a three-sphere (of radius $t$) and the line element is\begin{eqnarray}
    ds^2_{hyp} \approx dt^2 
-  \frac{t^2}{t_{0}^2} \left({dr'^2} + {r'}^2d{\Sigma}^2 \right) = ds^2_{R_h=ct}    
\end{eqnarray}

This is not accidental because, according to \citet{MCS2020}, the local conservative condition in dynamical systems only ensures internal consistency for $k=1$. Notice that the closed universe ($k=1$) implies that $\beta_0 = 0$, and the equation of the hypercone is homogeneous $\left|X - O \right|_{\eta} = 0$.

The new Friedmann equations can be obtained taking $\rho_{crit}(t) := {3{H}^{2}}/{8\pi G}  = \rho + \rho_{\Lambda}$, where $\rho$ is the density of the fluid of matter plus radiation, and ${\rho}_{\Lambda} := {\Lambda}/{8\pi G}$ is the dark energy density. With this, the hyperconical model leads to the following: \begin{eqnarray} \label{eq:rho.values}
4\pi \mathrm{G}\rho(1+w) = \frac{1}{t^2}
\\
\label{eq:lambda.values}
\Lambda t^2 = \frac{1+3w}{1+w}
\end{eqnarray} where $w$ is the parameter of the total equation of state. If $\Lambda$ is finite and constant, it is obtained that $\Lambda = 0$, G is constant, and $w = -1/3$ for any time
$t$ \citep{Monjo2017}. 


Globally, a map $f_{\gamma_0} : \mathbb{R}^{1,4}_{\eta}  \smallsetminus \lbrace O \rbrace \rightarrow  \mathbb{R}^{1,3}_g$ is required to remove the radial inhomogeneity produced in $\pazocal{H}^4$ by the comoving observes from $u$, satisfying $f = f_{g_0} \circ \mathcal{T}_{t}$  \begin{center}
	\begin{tikzpicture}[node distance=1.5cm, auto]
	\node (A) {$\pazocal{H}^4$};
	\node (B) [right of=A] {$\mathbb{R}^{1,4}_{\eta}$};
	\node (C) [below of=A] {$\mathbb{R}^{1,3}_g$};
	\draw[->](A) to node {$\mathcal{T}_{t}$}(B);
	\draw[->](A) to node [left] {$f$}(C);
	\draw[->](B) to node [right] {$f_{\gamma_0}$}(C); 
	\end{tikzpicture}
\end{center}For small regions, intrinsic  comoving distance $\hat{r}'$ can be defined by a $\alpha$-distorting stereographic projection,	
\begin{eqnarray} \label{eq:projectionmap}  
\hat{\gamma}':= f_{\gamma_0}(\gamma) :\approx \frac{\gamma}{\left(1-\frac{\gamma}{\gamma_0}\right)^\alpha}  \,,
\end{eqnarray} where $\gamma = \gamma(r') := \sin^{-1}(r'/t_0)$, $\gamma_0$ is the projective angle, and $\alpha = 0.5$ under symplectic symmetries \citep{MCS2023}. Locally, it is expected that $\gamma_0 \sim 2$, which is compatible with the fitted value of $\gamma_0 \approx 1.6^{+0.4}_{-0.2}$ when Type Ia SNe observations are used.
If curvature $k \neq 1$ is taking into account, the above equations need to replace $t_0 \to t_0/\sqrt{k}$ and allow for fitting $k$ as an additional parameter \citep{Monjo2017, Monjo2018}.

\subsection{Geometrical linkage: The fictitious acceleration} \label{sec:linkage}

When the comoving distance ($r'$) obtained from the hyperconical metric (Eq. \ref{eq:inhom0}) is projected with Eq. \ref{eq:projectionmap}, its radial inhomogeneity has the same behavior as an apparent acceleration \citep{Dam2017, Monjo2018, MCS2023}. To check this important feature, it is enough to analyze the Hubble parameter of the hyperconical model after the appropriate projection used. Of course, the resulting acceleration depends on the coordinate change (projection) used, but it is unique if the projection is locally a centered stereographic projection (see Appendix \ref{sec:accel}). For instance, taking the first-order expansion of the Hubble parameter of the flat-space standard ($H_{\Lambda CDM}$) and the hyperconical ($\hat{H}_{hyp}^{intr}$) model with $H_0 \equiv 1$, it is shown that \begin{eqnarray} \label{eq:hubble_lcdm00}  \nonumber
H_{\Lambda CDM\,}(z) &=& 
 \sqrt{\Omega_{r} +\Omega_{m} + \Omega_{\Lambda}}
+ \frac{4\Omega_{r} + 3\Omega_m}{\sqrt{\Omega_{r} + \Omega_{m} + \Omega_{\Lambda}}} \frac{z}{2} + O(z^2) 
\\ \nonumber
\label{eq:hubble_hyp0} 
  \hat{H}_{hyp\ }^{intr}(z) &=& \hspace{1mm} 1 \hspace{1mm}  + \hspace{1mm}  \frac{\gamma_0 - 1}{\gamma_0} z \hspace{1mm} + \hspace{1mm} O(z^2) 
\end{eqnarray}where the projective angle is locally $\gamma_0 \approx 2$. Moreover, simplifying with $\Omega_r \approx 0$, it is predicted that $\Omega_m \approx \frac{1}{3}$ and $\Omega_\Lambda \approx \frac{2}{3}$ under the first-order approach, which is compatible with estimation of $\Omega_m = 0.334 \pm 0.018$ and $\Omega_\Lambda = 0.666 \pm 0.018$ obtained by flat $\Lambda$CDM fitted to Pantheon+\&SH0ES \citep{Brout2022}. Thus, the deceleration parameter (Eq. \ref{eq:desac}) for $(\Omega_m, \Omega_\Lambda) = (1/3, 2/3)$ is \begin{eqnarray}
\label{eq:deceleration}
    q_0 := - \frac{\ddot a a}{{\dot a}^2}\,\bigg|_{t=t_0} \approx \frac{1}{2}\Omega_m - \Omega_\Lambda \approx - \frac{1}{2}
\end{eqnarray}

\begin{remark}[Acceleration is fictitious] In sum, linear expansion produces a radial inhomogeneity under dynamical embedding, and at the same time, this inhomogeneity mimics as an apparent acceleration compatible with the standard value. Thus, linear expansion leads to the value of $\Omega_{\Lambda} \approx 0.7$ (see \citep{MCS2023}).    
\end{remark}


\section{Some observational tests}
\label{sec:observational}

\subsection{Luminosity distance}

As a standard cosmological test, the luminosity distance of Type-Ia SNe was used to compare the performance of the standard and alternative models. Applying the Etherington's reciprocity relation $r_L = (1+z)\, {r'}$ \citep{Ellis2007} to the $\Lambda$CDM comoving distance $r'$, one can find:
\begin{eqnarray} \label{eq:comoving.distance} \nonumber   
  r_L^{{}^{\mathrm{\Lambda CDM}}} = \frac{1+z}{H_0}\,  \sin_K \int^{z}_{0} \frac{d{z'}}{\sqrt{ {\Omega}_{m} (1+{z'})^{3} + {\Omega}_{K} (1+{z'})^{2} + {\Omega}_{\Lambda}}} 
\end{eqnarray} where $\sin_K x := \lim_{\epsilon \rightarrow K} \epsilon^{-1/2} \sin (\epsilon^{1/2} x)$, that is, $\sin_0(x) = x$, $\sin_{+1} x = \sin x$, $\sin_{-1} x = \sinh x$, and recall that  $K = -\Omega_K H_0^2$ with $c\equiv 1$. The coasting and $R_h=ct$ models predict simpler relationships, \begin{eqnarray} \label{eq:lum1}
  r_L^{{}^{coast}} &=& \frac{1+z}{H_0}  \sin_K\left(  \ln(1+z)\right), \\  \label{eq:lum2}
    r_L^{{}^{R_h=ct}} &=& \frac{1+z}{H_0}\ln(1+z).
\end{eqnarray}  \nonumber  
Finally, for the hyperconical model, it is \citep{Monjo2017}: \begin{eqnarray} \label{eq:lum3}
   r_L^{{}^{hyp}} \;\; = \;\; \frac{1+z}{H_0} \, f_{\gamma_0}\left( \mathfrak{sn}_k \left( \ln(1 + z) \right)\right)
\end{eqnarray} where it is necessary to define the function
\begin{eqnarray} \label{eq:comoving.distance1} \nonumber
\hspace{1mm} \mathfrak{sn}_{k}^{-1} \left( \gamma \right)  \; := \,  \int_{0}^{\gamma} \frac{\sqrt{1-k^{-1}(1-\cos\gamma')^2}}{1-2k^{-1}(1-\cos\gamma')} {d \gamma'} \;=\; \gamma+O(\gamma^3)\end{eqnarray} with $\sin\gamma(r') :=  \frac{r'}{t_0} = H_0r'$.

Constraints on the model parameters were obtained using pairs of redshift and luminosity distance (modulus) from the Pantheon+ data of 1,701 light curves corresponding to 1,550 distinct Type Ia supernovae (SNe Ia) \citep{Scolnic2022, Brout2022}. The theoretical modulus, defined as \begin{eqnarray} \addtocounter{equation}{1} \label{eq:modulus} 
    \mu_{theo} := 5 \log (r_L/ \mathrm{Mpc}) + 25 \;= \; 5 \log (r_LH_0) + \tilde M   \,, \;\;
\end{eqnarray} has constant $\tilde{M}$ degenerate with $H_0$. The value of this constant can be obtained by minimization, in each case (\citealp[see for instance][]{Monjo2017}).

\begin{figure*} 
\centering
	\includegraphics[scale=0.8]{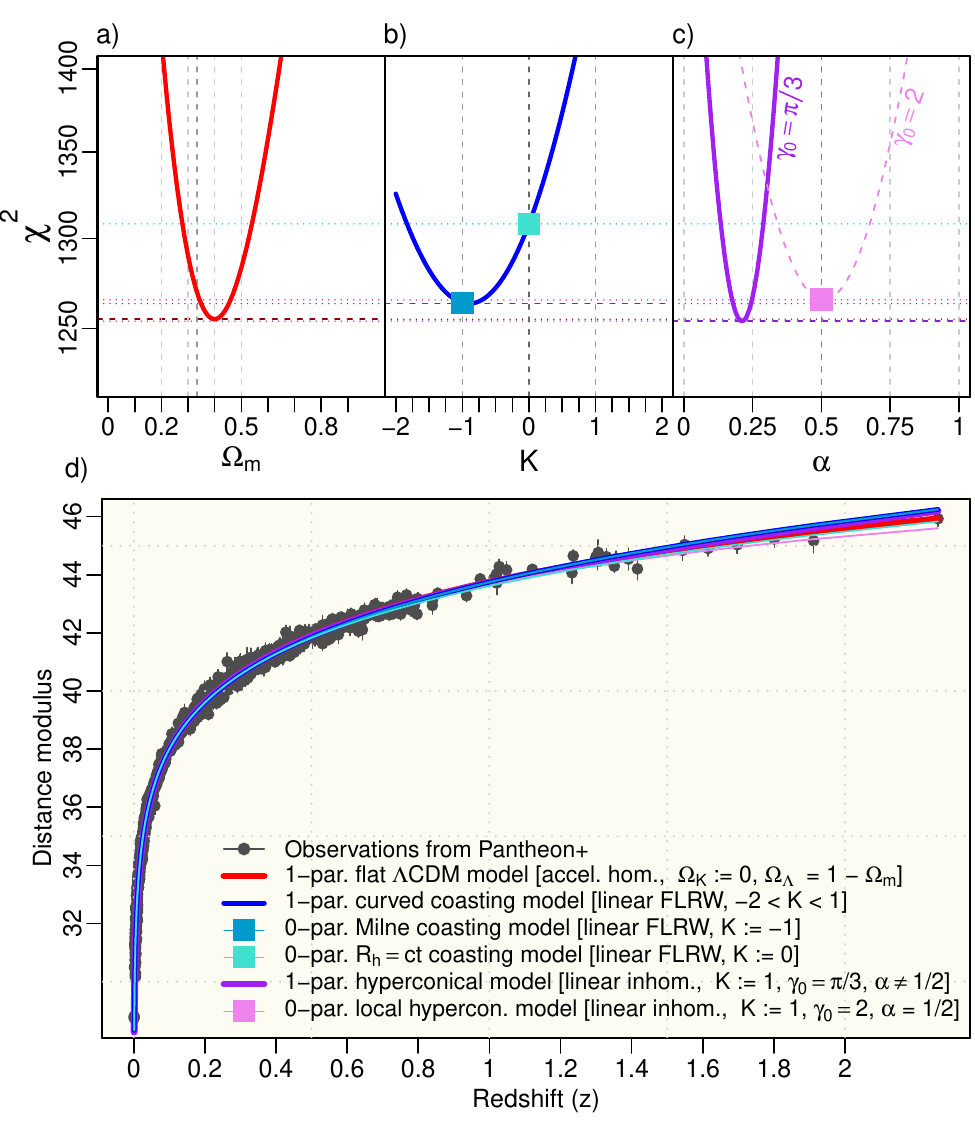}
    \caption{Model discrepancy ($\chi^2$) of theoretical distance modulus (Eq. \ref{eq:modulus}) constrained with Pantheon+ data for several cosmological models: a) fitting $\Omega_m$ of the standard $\Lambda$CDM model (red line), b) fitting curvature $K$ of the coasting model (blue line), compared to the prediction of the Milne model (sea blue square) and the $R_h=ct$ model (light blue square), c) fitting the distorting exponent $\alpha$ of the hyperconical model for $\gamma_0 = \pi/3$ (purple line) compared with the prediction of the local limit ($\gamma_0 = 2$ and $\alpha = \frac{1}{2}$, violet square), and d) final curves of the distance modulus as a function of redshift, according to the Pantheon+ data compared with the models. Error bars on the distance modulus are estimated from the diagonal of the covariance matrix.
}  \label{fig:Fig1}
\end{figure*}

 Finally, Pearson's chi-squared statistic $\chi^2 =  \sum_{i,j} \sigma_{{theo,i}} (\boldsymbol{\sigma}_{{obs}}^{-2})_{i,j}\sigma_{{theo,j}}$was considered as a measure of the discrepancy between each estimate of the model squared error $\sigma_{{theo,i}}^{2} := \left( \mu_{obs,i} - \mu_{theo,i} \right)^2$ and the observed error represented by the sample covariance matrix $(\boldsymbol{\sigma}_{{obs}}^{2})_{ij}$ and its inverse matrix $(\boldsymbol{\sigma}_{{obs}}^{-2})_{ij}$. In the case of Pantheon+, this matrix relates the covariance between SNe distance measurements due to various systematic uncertainties \citep{Scolnic2022, Brout2022}. The optimal theoretical modulus (Fig. \ref{fig:Fig1}d) was obtained by minimizing $\chi^2$ for each model.



\begin{remark}[Indistinguishable for SNe Ia data] The results of the fitting pointed to a good performance of all the models, slightly favoring standard $\Lambda\mathrm{CDM}$ and hyperconical models, with $\chi^2 = 1255$ for both cases, with the parameters $\Omega_m = 0.4 \pm 0.1$ ($\Omega_K \equiv 0$) and $\alpha = 0.24 \pm 0.06$ ($\gamma_0 \equiv \frac{\pi}{3}$). However, the local prediction of the hyperconical model ($\gamma_0 \approx 2$, $\alpha = 0.5$) obtained $\chi^2 = 1265$ (Fig.~\ref{fig:Fig1}). The coasting model was fitted with $\chi^2 = 1,264$ for $k_{coast} \approx -1.0$, which corresponds to the Milne model, while $R_h=ct$ obtained $\chi^2 = 1308$ without any free parameter. However, the apparently modest result of the $R_h=ct$ model is noticeably improved if empirical fitting of the SN Ia light curves (SiFTO) is used \citep{Conley_2008}. Using the Bayes information criterion, \citet{Wei_2015} found that $Rh = ct$ is preferred over the standard model with a probability of $\sim$ 90\% compared to $\sim$ 10\%.
\end{remark}

\subsection{Angular diameter}

Another standard rule to validate a cosmological model is the \textit{angular diameter distance}, $d_A$, which is defined by the relation between the physical size $\ell$ of an object and its angular size $\theta$, as viewed from an instrument\begin{eqnarray}
     d_{A}={\frac{\ell}{\theta }}\,.
\end{eqnarray}Theoretical predictions of the standard, coasting and hyperconical models were contrasted by comparing them to model-independent observations collected from galaxy clusters \citep{DeFilippis_2005} and quasars \citep{Cao2017}. The angular diameter distance of a cosmological model is directly given by the comoving distance $r'$ as in the following:\begin{eqnarray} 
    d_A (z) = \frac{r'(z)}{(1+z)} = \frac{r_L(z)}{(1+z)^2}
\end{eqnarray}where $r_L(z)$ can be taken from Eqs. \ref{eq:lum1}--\ref{eq:lum3}. Because of the geometrical features of the metric used, the maximum angular distance can be reached at a finite redshift $z_{\max} \in (0, \infty)$.

\begin{figure*}
\centering
	\includegraphics[scale=0.78]{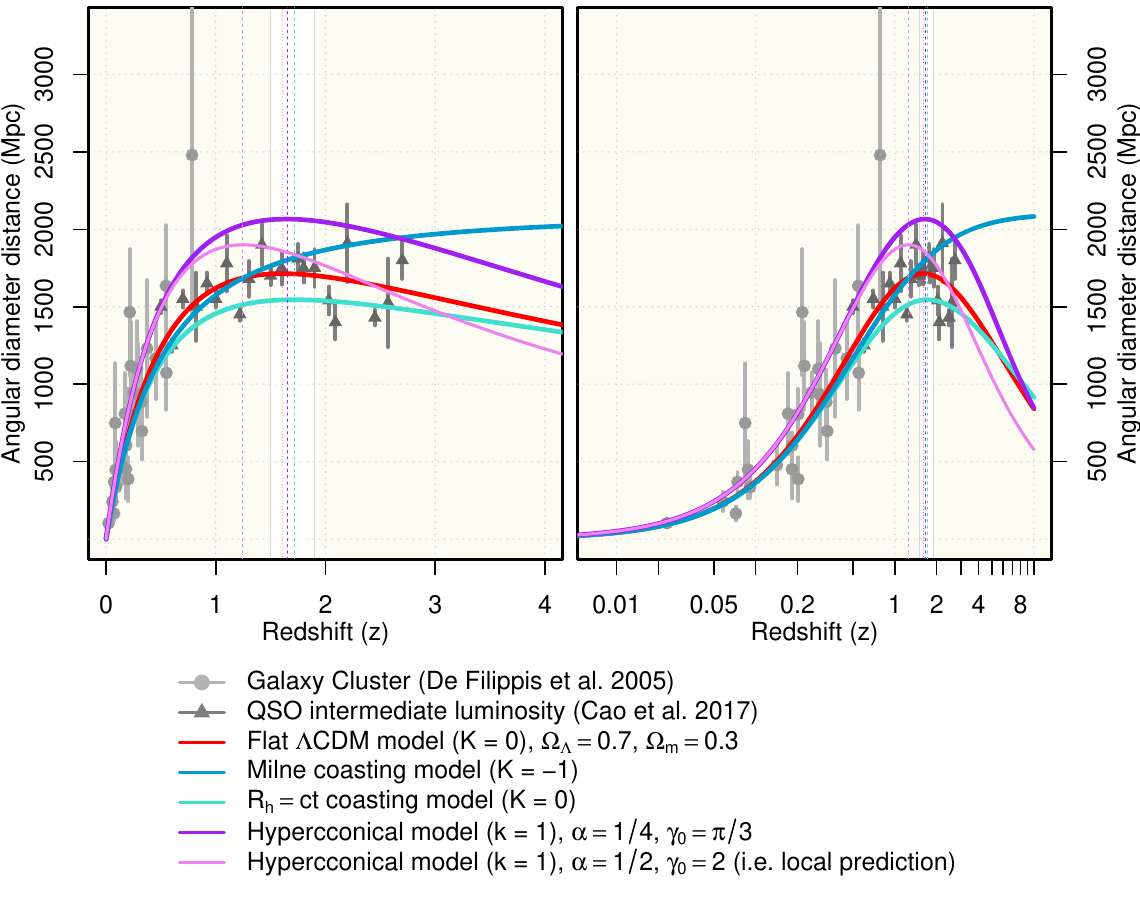}
    \caption{Theoretical predictions of angular diameter distances $d_A(z)$ according to different cosmological models and comparison with model-independent observations collected from 25 galaxy clusters \citep{DeFilippis_2005} and 21 samples gathered from 107 quasar intermediate luminosity \citep{Cao2017}. The white vertical windows correspond to the $1.5 \le z \le 1.9$ preferred range for 140 compact quasar cores \citep{Melia2018b}. The dotted vertical lines represent the location of the maximum angular diameter distance for each model.}
    \label{fig:Fig2}
\end{figure*}

\begin{remark}[Similar model-independent sizes] Accelerated and linear cosmological universes predict an angular diameter distance compatible with the model-independent observations of galaxy clusters and quasar intermediate luminosity (Fig. \ref{fig:Fig2}). Only the Milne model ($K = -1$) can be excluded because its prediction is $z_{\max}^{Milne} = \infty$. \citet{Melia2018} found $z_{\max}^{obs} = 1.7 \pm 0.2$ by employing Gaussian processes in a sample of 140 objects. \end{remark}

For the standard and $R_h=ct$ models, the maximum angular diameter distance is found in $z_{\max}^{\Lambda cdm} \approx 1.59$ and $z_{\max}^{R_h=ct} \approx 1.72$, while the hyperconical model leads to $z_{\max}^{Hyperc} \approx 1.65$. When combining the luminosity and angular distances, the most favored models are the standard $\Lambda$CDM, the $R_h=ct$ and the hyperconical models. As expected, the local version ($K=1$, $\alpha = \frac{1}{2}$, $\gamma_0 = 2$) of the hyperconical model only predicts well for low redshift values ($z < 1$) because its maximum is at $z \approx 1.25$.

\subsection{Cosmic chronometers and BAO}
\label{sec:cc}

As a third observational test, the redshift dependency of the Hubble parameter was contrasted with the accelerated (standard) and linear (coasting) cosmological models. Specifically, a total of 51 estimates of the Hubble parameter were considered (Table \ref{tab:B1} of Appendix \ref{sec:dataCC}) according to cosmic chronometers (CC: 34), Type-Ia SNe distance ladder (Ladder: 3), and three different data sets of BAO signals: Radial BAO size in galaxy distribution (BAO-Gal: 7) \citep{Gaztañaga2009, Blake2012}, redshift space distortions (RDS) from Baryon Oscillation Spectroscopic Survey (eBOSS) quasars (BAO-QSO: 4) \citep{Zhao2018} and Lyman-alpha forest emission in quasar spectra (BAO-Ly$\alpha$: 3) \citep{FontRibera2014, Delubac2015, Dong2022}. 

\begin{figure}  
\centering
	\includegraphics[width=140mm]{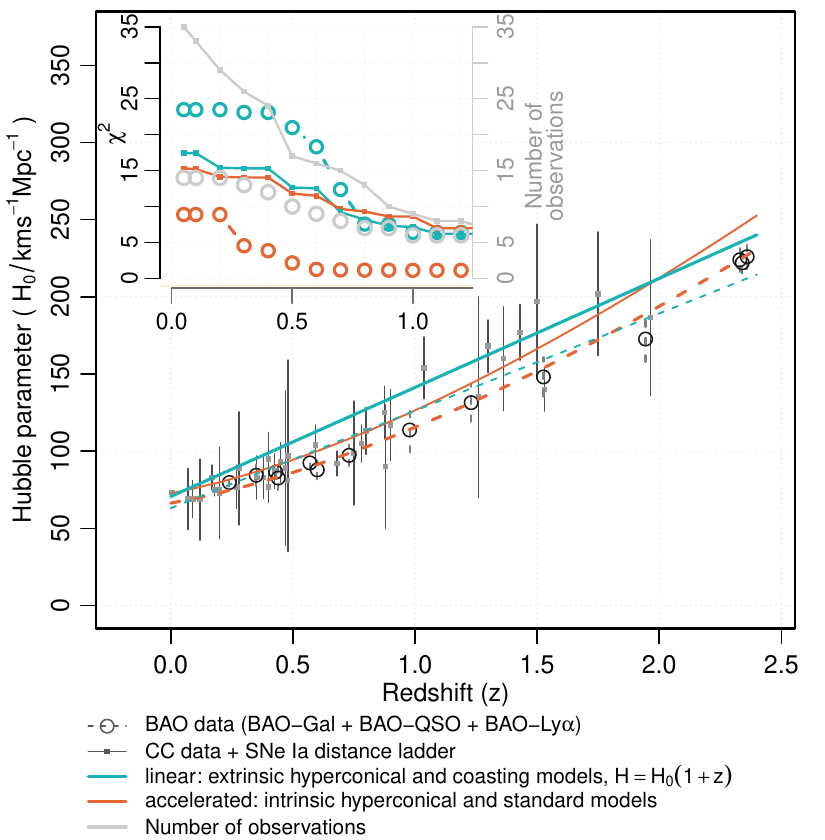}
    \caption{Fits of the linear (extrinsic hyperconical/coasting) and accelerated (intrinsic hyperconical/standard) models to the observations of the Hubble parameter estimated from cosmic chronometers and radial BAO size methods in galaxy distribution (Gal), quasars (QSO), and Lyman-$\alpha$ forest (Ly-$\alpha$), respectively. The inner sub-figure shows the Pearson chi-squared statistics ($\chi^2$) obtained for the fits restricted by $z > z_0$ with respect to a particular threshold $z_0$.
} \label{fig:Fig3}
\end{figure}

Taking into account CC and the distance ladder of type Ia SNe data, the Hubble parameter fitted by the hyperconical model ranges between $h = 70.6 \pm 1.2$ under the extrinsic framework and $72.0 \pm 1.3$ for the \textit{intrinsic} model. However, the \textit{intrinsic} value of Hubble is reduced to $h = 66.33 \pm 0.65$ if the radial BAO size methods are considered (Fig. \ref{fig:Fig3}). The physical interpretation is that the cosmological timeline is better described by CC and purely geometrical techniques in SNe data, leading to the unitary relationship of $Ht \equiv 1$ (e.g. $t = 13.80 Gy \to h = 70.86$; \citep{MCS2023}). However, if the standard accelerated-based model is forced to be adjusted to purely geometrical observations, the fitted value is $72.0 \pm 1.3$. \citet{Wei2020} found a compatible value of $h ={75.3}_{-2.9}^{+3.0}$ by using the time delay measurements of seven strong lensing systems and the known ultraviolet versus X-ray luminosity correlation of quasars. These high values are in tension with respect to the estimation of $h = 66.33 \pm 0.65$ obtained with BAO data, the $h = 67.4 \pm 0.5$ measured in CMB by Planck \citep{Planck2020a}, and the $h < 70$ estimated with old astrophysical objects \citep{Wei2022}. 

\begin{remark}[Hubble discrepancy] Different values of the Hubble parameter would be due to the different geometrical perspectives assumed.
\end{remark}

To alleviate the Hubble tension, the hyperconical model proposes that the highest values of the Hubble parameters correspond to ``extrinsic estimations'' \citep{MCS2023} because they are linked to cosmology-independent measurements such as purely geometrical methods (e.g., CC or distance ladder). On the other hand, the lowest values are related to model-dependent light paths, which are proper distances from an intrinsic viewpoint. However, the results and their interpretation are very sensitive to the type of data used. Consistently, \citep{Wei_2017} showed that cosmological tests are more robust when only a truly model-independent data set (e.g., CC) of $H(z)$ is used, without mixing with high locally measured values of the Hubble parameter $H_0$. In other words, the CC data seem to point to a constant expansion rate rather than a \textit{Hubble bubble} \citep{Wei_2017, Wu2017}. 

The deceleration parameter (\ref{eq:deceleration}) predicted by the intrinsic viewpoint of the hyperconical model is $q_0 = -1/2$, but under purely geometric approaches, it is expected to be $q = 0$ for all redshifts. By applying cosmographic techniques to high-redshift ($z>1$) data, \citet{Pourojaghi2022} found a deceleration of $q_0 = -0.24\pm 0.42$ for SNIa and $q_0 = -0.095^{+0.49}_{-0.28}$ for QSO, which is $q_0 = -0.16^{+0.45}_{-0.34}$ for the combination of both samples in high-redshift regions. Similar results were found by \citep{Colgain2022} when QSO samples are extended up to $z = 1$, with $q_0 =  0.0755^{+0.2760}_{-0.3465}$ (from a flat $\Lambda$CDM model with $\Omega_m = 0.717^{+0.184}_{-0.231}$).


\section{Paradigm shift in GR}
\label{sec:refoundation}

\subsection{Cosmological principles and zero active mass}

The current standard model of modern cosmology has received some critiques because of its auxiliary unobserved quantities (e.g., dark matter and dark energy) as a ``conventionalist stratagem'' in response to other unexpected observations \citep{kroupa_2012, Merritt2017}. Even the cosmological principle could be questioned by these observations \citep{Aluri2023}. And the derivation of the FLRW metric is clear in the spatial section, but it is arbitrary to set the temporal component \citep{Monjo2017, Monjo2018, Melia2022}.

The abundance of galaxies is satisfactorily modeled by standard cosmology for low-redshift objects ($z < 10$) \citep{Behroozi2019}. However, the discovery of large objects at $z>10$, the so-called ``impossible early galaxies'', poses serious tensions in $\Lambda$CDM cosmology because they are earlier than expected \citep{Casey2023}. Alternatively, MOND-based theories suggest a higher growth rate than predicted by the standard model \citep{Sanders1998,Sanders2008,Wittenburg_2020,Wittenburg_2023}. This is compatible with linear expansion, since the hyperconical model allows one to formulate a relativistic theory with MOND regimes, thus explaining the well-known acceleration--mass discrepancy and the baryonic Tully-Fisher relation (BTFR, which is observed in galaxy rotation curves) \citep{Chardin2021, Monjo2023}. Nevertheless, according to \citep{John2023}, linear expanding (coasting) universes are sufficient on their own to allow galaxy formation at redshift up to $z \sim 20$, which corresponds to $\sim 700$ Myr.   Furthermore, linear expansion may better explain most of the flagship cosmological observations than the standard model \citep{Melia2018}.

Therefore, the existence of exotic dark matter and dark energy would not be necessary to explain the expansion or the rotation of galaxies \citep{kroupa_2012, kroupa_2013}. The non-detection or evidence of dark matter is clear in the solar system, indicating that dark matter would not interact with anything other than itself \citep{Gaidau_2019}. According to \citet{Belbruno2022}, the dark matter effect should be considered for objects when leaving the solar system. However, \citet{Brown_2023} found that MOND provides an accurate explanation for the anomalies of Kuiper Belt objects that lie beyond Neptune, without the need for dark matter. Since excess gravity seems to be related to the amount of visible baryons (according to the BTFR), the simultaneous statements that ``dark matter does not interact with baryonic matter'' and ``dark matter is tightly bound to baryonic matter'' is a contradiction. The alternative cosmology is the simpler $R_h=ct$ perspective of \citet{Melia2018}, which can be derived from the hyperconical framework, as this also produces MOND behavior with galaxy rotation diversity \citep{Monjo2023}. This is an advantage because MOND-like models can explain other observations such as the KBC void and the dynamics of wide binary stars, among others \citep{Haslbauer2020, Chardin2021, Chae_2023, Chae_2023b}.

However, strong criticisms were made regarding the ``zero active gravitational mass'' of the linear model \citep{Mitra2014}. The solution proposed by the hyperconical paradigm is the modification of the Lagrangian density to make compatible a background metric independent of the total matter content. It is equivalent to assume that GR is valid only at local scales, so the total Lagrangian density modifies the standard Einstein--Hilbert term, as follows \citep{MCS2020, MCS2023, Monjo2023}:
\begin{eqnarray} \addtocounter{equation}{1} \label{eq:space.lagrangian}
\mathcal{L} = \frac{1}{16\pi\mathrm{G}} \Delta R +
\mathcal{L}_M 
\end{eqnarray} where $\Delta R := R - {\tilde R}$ is the scalar curvature perturbation with respect to the local Ricci curvature of the background hyperconical metric, ${\tilde R} = -{6}/{t^2} = -\rho_{crit}$, while $\mathcal{L}_M = -\rho_M$ is the Lagrangian density of mass--energy that contributes to $R$ (i.e., it is any local gravity system). With this, GR is consistent with the Melia's definition of 'zero active gravitational mass' (equation of state $w = -1/3$). Therefore, the only difference between this proposed modification and the standard GR is the transitional large scales (low accelerations), which would produce MOND-like regimes from fictitious cosmic acceleration \citep{Monjo2023}, allowing the theory to be falsified.

To better understand how this issue ('zero active mass') is linked to the cosmological principles, let us approach the universe as an (almost) isotropic and homogeneous distribution of point masses through space (see, e.g., Ref. \citep{Fleury2017}). Then, because of homogeneity and isotropy, Gauss's law ensures that the total gravitational force is zero at every point mass and at the mid (Lagrange) points, but not in its surroundings. Therefore, a homogeneous distribution of discrete masses does not effectively contribute to the dynamics of the background metric but only to the local behavior. The closer to the continuum, the closer to zero the gravity field produced by the distribution of masses will be. This result is similar to that obtained inside an infinite plane with homogeneous electrical charge density, which only presents nonzero electrical field outside the plane, in its perpendicular direction. Therefore, the replacement of $R \to \Delta R$ is essential to determine that mass perturbs spacetime but it does not impact on shape or evolution of the universe.

\subsection{Local general relativity and Schwarzschild-like metric}

According to the new paradigm, matter locally perturbs the metric, but the global background metric is independent of the matter content, because it is inherited from embedding in the (Lorentzian) ambient spacetime. Now, let ${\tilde g}$ be the background metric, corresponding to the hyperconical universe, and let $g := {\tilde g} + h$ be the total metric with perturbation $h$. The background Ricci curvature tensor and its scalar are \citep{MCS2020}: \begin{eqnarray}
    {\tilde R}_{00} = 0,\;\; {\tilde R}_{0i} = 0,\;\; {\tilde R}_{ij}  = -\frac{1}{6}(g^{kl}\partial_t {\tilde g}_{kl})^2 {\tilde g}_{ij} = -\frac{2}{t^2}{\tilde g}_{ij}\;\;\;\;\;\;\;\;\;
    \\
    {\tilde R} =  -\frac{1}{2}({\tilde g}^{kl}\partial_t {\tilde g}_{kl})^2 = - \frac{6}{t^2}\,,\;\;\;\;\;\;\;\;\;
\end{eqnarray}while total Ricci scalar is $R = g^{\mu\nu}R_{\mu\nu} = \tilde R + \Delta R$, where the perturbation scalar $\Delta R$ logically satisfies the Einstein field equations according to Eq. \ref{eq:space.lagrangian}; that is, $\Delta R_{\mu\nu} - \frac{1}{2}\Delta R g_{\mu\nu} = 8\pi\mathrm{G}T_{\mu\nu}$.  Therefore, the new equations for the total contribution $R$ are\begin{eqnarray}
    R_{00} - \frac{1}{2}R\,g_{00} -  \frac{3}{t^2}g_{00} = 8\pi\mathrm{G}T_{00} \;\;\;\;\;\;\;\;\;
    \\
    R_{ij} - \frac{1}{2}R\,g_{ij} - \frac{1}{t^2}g_{ij} =  8\pi\mathrm{G}T_{ij}\,, \;\;\;\;\;\;\;\;\;
\end{eqnarray}where $T_{\mu\nu}$ is the stress-energy tensor. This is equivalent to consider a ``cosmological (almost) constant'', or dark energy with equation of state $w = -1/3$, as mentioned in the previous section. 

The above modifications of the Einstein-Hilbert Lagrangian and of the Einstein field equations are minimum at the local scale, but they are fundamental to make compatible with the linear expansion of the metric. However, a fictitious acceleration emerges when projected coordinates are used (see Sec. \ref{sec:linkage}), contributing to an additional term in the gravity \citep{Monjo2023}.

Without considering the projection under the linear expansion, the Schwarzschild (vacuum) solution has an approximated equivalence. To check this feature, let $k = 1/(\nu t_0)^2
$ be the spatial curvature with $\nu^2 := 1-\beta_0^2$. Then, the background metric of the hyperconical universe (Eq. \ref{eq:inhom0}) can be approximated up to second order of $r'^2/(\nu^2t_0^2)$ as follows: \begin{eqnarray}
ds^2_{hyp} &\approx& dt^2 \left(1- \frac{r'^2}{t_0^2} \right) 
- 
\addtocounter{equation}{0} \label{eq:inhom0b}
\frac{t^2}{t_{0}^2} \left[\left(1+\frac{r'^2}{\nu^2 t_0^2}\right)dr'^2 + {r'}^2d{\Sigma}^2 \right] -  \;\; \frac{ 2r't}{t_{0}^2}dtdr'  \;\;\;\;\;\;\;\;
\end{eqnarray}where the spatial term $r'^2/(t_{0})^2$ as a basis of critical density $\rho := 3/(8\pi\mathrm{G}t^2)$. Therefore, taking into account the local conservative condition $k \equiv 1 \equiv \nu$, or $\beta_0 = 0$, derived from dynamical systems \citep{MCS2020}, the spatial term is \begin{eqnarray}\label{eq:metric_M}
\frac{r'^2}{t_{0}^2}   =  \frac{r^2}{t^2}= \frac{2\mathrm{G}\rho \frac{4}{3}\pi r^3}{r}  = \frac{2\mathrm{G}\mathcal{M}(r)}{r} \,,
\end{eqnarray}with total energy $\mathcal{M}(r) := \rho \frac{4}{3}\pi r^3$. Adding a perturbation in density such as $\rho \to \rho + \Delta\rho(r)$, Eq. \ref{eq:metric_M} is now:  \begin{eqnarray} \label{eq:metric_per}
\frac{{r'}^2}{t_{0}^2} \to \frac{2\mathrm{G}(\rho+\Delta\rho(r))\frac{4}{3}\pi r^3}{r}  = \frac{r'^2}{t_{0}^2} + \frac{2\mathrm{G}m(r)}{r} \,,
\end{eqnarray} where $m(r) := \rho(r)\frac{4}{3}\pi r^3$ is the energy of the perturbation. Finally, total metric $g_{\mu\nu}$ can be obtained by replacing Eq. \ref{eq:metric_per} in Eq. \ref{eq:inhom0b}, and considering $r' << t_0$, it is obtained a Schwarzschild-like metric:\begin{eqnarray} \nonumber
  & d&s^2_{Sch}  := g_{\mu\nu}\big|_{r' << t_0} dx^\mu dx^\nu  \approx
  \\ \addtocounter{equation}{1} \label{eq:Schwarzschild_g}
  & & \left(1- \frac{2\mathrm{G}{M}}{r} \right) dt^2
-  \frac{t^2}{t_{0}^2} \left[\left(1 + \frac{2\mathrm{G}{M}}{r}\right) dr'^2 + r'^2d{\Sigma}^2 \right]\,.\;\;\;\;\;\;
\end{eqnarray} which is also $ds^2_{Sch} \approx (\eta_{\mu\nu} + h_{\mu\nu})dx^\mu dx^\nu$, where $h_{\mu\nu} := g_{\mu\nu} - \tilde g_{\mu\nu}$. Notice that, for small gravitational systems, one can assume that $t \approx t_0$ and $r' \approx r$.

Nevertheless, for large gravitational structures, the distorting stereographic projection need to be considered, and the perturbation becames as follows \citep{Monjo2023}:\begin{eqnarray} \nonumber \label{eq:gtt}
h_{00} &\approx&   - \frac{2\mathrm{G} M}{r}  +  \frac{2}{\gamma_0} \frac{r}{t}  
\end{eqnarray}  where $\gamma_0 \ge 2$ is a parameter that produces a MOND-like regime, depending on the geometry of the gravitational system considered. For example, for galaxy rotation curves, it is $\gamma_0 \approx 10$.

\subsection{Apparent flatness and zero active mass}

As detailed in \ref{sec:linkage}, the intrinsic viewpoint of the hyperconical universe leads to a fictitious acceleration (represented by a kind of dark energy $\Omega_{\Lambda} \approx 0.7$) that compensates for the expected gravity of an estimated amount of matter ($\Omega_m \approx 0.3$, mostly dark matter) to obtain a total density equal to the critical density. This implies that the metric should be spatially flat with a balance of forces that lead to linear expansion (i.e., zero active mass).

\begin{remark}[\textbf{The background metric of the universe does not depend on the content of matter}] Linear expansion is directly inherited from minimal embedding in a 5-dimensional Minkowski space. However, the standard model is based on the validity of GR for cosmological scales, which establishes that spacetime curvature is intrinsically determined by the matter content. According to the FLRW family of metrics, the spatial curvature of the universe is positive (closed) if the universe is dominated by matter, negative (open) if the universe is dominated by the expansion mechanism (whatever it is), and flat (or critical) if there exists a perfect equilibrium between the matter-related gravity to be closed and the expansion-based pressure to be open. If the universe behaves as a flat-space manifold, this means that gravity and expansion pressure are perfectly balanced to avoid that matter content impacts the metric (notice that it is only flat in space since the Ricci scalar $R \approx -6\dot{a}^2/a$ is not zero due to the contribution of the expansion scale factor $a$). In other words, an ``apparently flat space'' implies a zero active gravitational mass-energy and it also implies a linear expansion.
\end{remark}




\section{Concluding remarks: Conciliation of perspectives}
\label{sec:perspectives}

Recent cosmological observations have forced the standard model to address new challenges with additional parameters. A simpler alternative, the so-called $R_h=ct$ model, is a competitor to explaining most of the current issues. Apparently, the standard accelerated and linear-expanding $R_h=ct$ universes are incompatible. However, another (geometrically based) model provides a bridge to connect both perspectives: the hyperconical model is built from the dynamic embedding of homogeneous and isotropic four-dimensional manifolds into a flat five-dimensional Minkowskian spacetime. Then, the linear expansion naturally emerges from this embedding, and the hyperconical manifold has a locally flat space, recovering the $R_{h}=ct$ model. On the other hand, setting linearly comoving observers as the references, a radial inhomogeneity appears in the metric (representing the lapse and shift terms). However, the coordinates can be projected onto a four-dimensional \textit{accelerated hyperplane} by using a distorting stereographic projection that assimilates the radial inhomogeneity as a fictitious acceleration compatible with the standard model. Thus, non-standard particles of dark energy and cold dark matter would be unnecessary, and most observational tensions would be automatically alleviated \citep{MCS2023, Asencio_2023}.


Therefore, the hyperconical paradigm proposes two different geometrical perspectives: i) the extrinsically homogeneous linearly expanding hypercone equivalent to the $R_h=ct$ model at every point and ii) the intrinsically apparently accelerated flat $\Lambda$CDM. Schematically, the two perspectives can be summarized by the key terms as follows (a brief explanation of each term is marked in gray): \begin{eqnarray} \nonumber
\resizebox{0.71\hsize}{!}{
$
\begin{matrix}
\mathrm{Hyperconical} \\
\mathrm{ model} \\
\textcolor{gray}{\mathrm{(closed,}} \\
\textcolor{gray}{\mathrm{linear\; \&}} \\
\textcolor{gray}{\mathrm{homogeneous)}}
\end{matrix}
 \Rightarrow
    \begin{cases}
    \begin{matrix}
\mathrm{Local} \\
\mathrm{extrinsic} \\
\mathrm{viewpoint} 
\end{matrix}
\;\equiv\;
\begin{matrix}
R_h=ct \; \mathrm{metric}\\
\textcolor{gray}{\mathrm{(flat,\; linear\; \&}} \\
\textcolor{gray}{\mathrm{homogeneous)}} \\
{}
\end{matrix}
\\
\begin{matrix}
\mathrm{Local} \\
\mathrm{intrinsic} \\
\mathrm{viewpoint} 
\end{matrix}
 \;\equiv
 \begin{matrix}
\mathrm{Radial\; inhom.} \\
\mathrm{metric} \\
\textcolor{gray}{\mathrm{(closed,}} \\
\textcolor{gray}{\mathrm{linear\; \&}} \\
\textcolor{gray}{\mathrm{inhomogeneous)}}
\end{matrix}
  \equiv\;  
  \begin{matrix}
\mathrm{Apparent } \\
\Lambda\mathrm{CDM \; model} \\
\textcolor{gray}{\mathrm{(flat,}}\\
\textcolor{gray}{\mathrm{``accelerated''\;\&}}\\
\textcolor{gray}{\mathrm{homogeneous)}} \\
+ \\
\mathrm{MOND}
\end{matrix} 
    \end{cases} \hspace{1mm}
    $%
    }
\end{eqnarray}
    
The hyperconical universe is extrinsically closed (curvature $k \equiv k_{hyp} = 1$), with a radius equal to the age $t$ of the universe, expanding at the speed of light (i.e., $a = t/t_0$) and with a locally flat spatial curvature ($k_{hyp} = 1 \Leftrightarrow K_{FLRW} = 0$), as required in the $R_h = ct$ model. More specifically, the Ricci scalar curvature is equal to that obtained for a flat FLRW metric with linear expansion.

The reviewed observational tests lead to similar conclusions concerning the double perspective. On the one hand, the linear cosmic timeline of the coasting/$R_h=ct$ models is favored by the so-called ``impossible early galaxies'' and maximum angular diameter distance, as well as other model-independent observations such as the well-known cosmic chronometers and the purely geometrical methods (e.g., ladder distance). On the other hand, the model-sensitive luminosity distance seems to favor the standard model because it is estimated by a physical dependency of the light paths. However, \citet{Wei_2015} showed that SiFTO favors $R_h=ct$ over the $\Lambda$CDM model for Type Ia supernovae data. 

The only remaining critique for linear expansion models is that they are non-physical because are empty or have an exotic equation of state $w=-1/3$. However, that critique is performed by assuming that GR is valid for cosmological scales. The hyperconical paradigm proposes a modified gravity Lagrangian density. With this proposal, GR is directly valid at local scales, but it needs to extract the cosmological Ricci scalar ${\tilde R} = -6/t^2$ as a boundary condition because zero active gravitational mass is derived from linear expansion. In other words, the geometry and evolution of the hyperconical universe are independent of the matter content.

The interpretation of the additional scalar curvature term (${\tilde R} = -6/t^2$) is that it is an ``integration constant'' that depends on the age of the universe, similar to a dynamical dark energy. Therefore, the ``constant'' proposed by Einstein to keep the universe unaltered (static) by the matter content is now reinterpreted as a boundary condition to keep the universe with inertial expansion (inherited from the ambient manifold). General relativity should be employed as a perturbation theory of the background metric of the universe (see details in \citep{Monjo2023}), which is determined by the difference between the total density energy $\rho$ and the background density $\rho_{crit}$: Energy perturbations $\Delta \rho = \rho - \rho_{crit} =: \rho_M$ contribute to the Ricci scalar $R_M := \Delta R = R - {\tilde R}$.

In summary, we propose a new paradigm to understand the role of general relativity in cosmology as a perturbation theory, and we claim a conciliation between observations by interpreting a double perspective of linear expansion and the fictitious acceleration forced by comoving observers \citep{MCS2023}. Galaxy rotation curves are well modeled by the proposed model \citep{Monjo2023}, and therefore exotic dark-matter particles are unnecessary. Moreover, linear expansion removes the horizon and flatness problems, and thus dark energy and inflation are not required.  

\smallskip
  \smallskip

\section*{Acknowledgements}

The author thanks the very helpful review and contributions of Prof. M. John. Also, it is necessary to acknowledge the very valuable comments of Prof. Rocky Kolb, Prof. Fulvio Melia, and Prof. Jun-Jie Wei. Finally, the author greatly appreciates the anonymous reviewers for their many insightful comments and suggestions.

\smallskip
  \smallskip
  
\section*{Data Availability}

In this study, no new data was created or measured.

\smallskip
  \smallskip
  

\appendix

\section{Minimal dynamical embedding}
\label{sec:embedding}

\subsection{What is it?}

This appendix develops the derivation of the hyperconical metric. It includes the proof that the minimal dynamical embedding of a homogeneous and isotropic universe is the hyperconical metric, that is, a linearly expanding universe with an apparent radial inhomogeneity, whose Ricci scalar is locally the same as the one obtained from a spatially flat FLRW metric with linear expansion. 

\subsection{Static and dynamical embeddings}

Classically, FLRW metrics are derived in two steps: the first step is to obtain the spatial part of the metric from a static embedding in a flat four-dimensional Euclidean space, and the second step is to introduce a time-dependent scale factor $a(t)$ and the temporal coordinate $t$. However, this procedure is not adequate for (dynamical) Lorentzian manifolds because time and space are mixed in the same space-time continuous. 

 Let $\vec{\ell}$ be the spatial coordinates of an expanding manifold, so they can be rewritten using comoving coordinates $\vec{\ell}'$ and a scaling factor $a(t)$ varying with time $t$.
For instance, let $S_{R}^{3} := \left\{ \vec{\ell} \in \mathbb{R}^{4} : |\vec{\ell}| = R(t) \right\}$ be an expanding 3-sphere of radius $R(t) \in \mathbb{R}_{\geq 0}$ and centered in the origin, but the reasoning is equivalent for the symmetric 3-hyperboloid space. Then, the scaling factor is $a(t) = {R(t)}/{R(t_0)}$ and total spatial coordinates are $\vec{\ell} = a(t){\vec{\ell}}' \in S_R^{3}$. Therefore, the differential lines contains both spatial and temporal contributions, that is, $d\ell = \textit{a}(\textit{t})d\ell' + \ell' da(t)$. Notice that the second term can be assimilated in the spatial term as a radial inhomogeneity \citep{Monjo2017, Monjo2018}. 

\subsection{Minimal singular manifold}

In the 5-Minkowskian spacetime, $\mathbb{R}_\eta^{1,4} := (\mathbb{R}^5, \eta_{1,4})$, the maximum homogeneous and isotropic Lorentzian manifolds can be spatially flat, hyperboloid and hypersphere. However, the unique homogeneous manifold without boundary that has finite volume is the hypersphere. 

\smallskip

\begin{proposition}[Regular dynamical manifolds are closed] \label{prop:minsingular} If $M \in \mathbb{R}_\eta^{1,4}$ is a minimal singular (or maximal regular) four-dimensional hypersurface with finite parameter time, it also has finite spatial volume, so it is closed.

\textbf{Proof}. As a regular (smooth) hypersurface, $M$ can be parameterized by a coordinate system. Moreover, as a dynamical system, its evolution depends on a time-like coordinate. Because $M$ is maximal regular, it is globally hyperbolic, foliated by Cauchy hypersurfaces, and there exists a causality parameterization by ``proper time'' $t$, like an arc length, for which the evolution speed $c$ of the Cauchy hypersurfaces is unity, $c \equiv 1$. Because the timeline of $M$ is finite by hypothesis, there exist two finite times $t_{-}, t_{+} \in \mathbb{R}$ for which $M(t)$ is not defined (does not exist) at $t_{-}$ but is at $t_{+}$, and there also exists a time $t_0 \in (t_{-}, t_{+})$ for which the $\mathrm{Volume}(M(t)) < \epsilon$ approaches 0 by minimal singularity hypothesis (excluding 0). Therefore, the finite $c$ and the finite $t$ lead to a finite volume, $0 < \mathrm{Volume}(M(t)) < \infty$.  \ensuremath{\blacksquare}   
\end{proposition}

\smallskip

Therefore, only the expanding hyperspheres $S_t^3$ are maximal regular objects embedded in Lorentzian (dynamical) manifolds. Moreover, taking $t$ as a temporal parameter and the unitary velocity ($c \equiv 1$) by regular parameterization, the foliation of these hyperspheres is built by a hypercone \begin{eqnarray} \nonumber
\pazocal{H}^4 \; :=  \; S_{\mathbb{R}_{>0}}^3 \; :=  \;\left\{ X  \; \in  \; \mathbb{R}_{>0} \times \mathbb{R}^4 \subset \mathbb{R}_\eta^{1,4} \; : \; \left|X \right|_{\eta_{1,4}} = 0  \right\}. \\  
{}   \addtocounter{equation}{1} {} 
\label{eq:H}
\end{eqnarray} In Minkowskian coordinates, $(t, x, y, z, u) =: (t, \vec{r}, u)$, the equation of the hypercone is $0 = \left|X \right|_{\eta_{1,4}} = t^2 - x^2 - y^2 - z^2 - u^2 = t^2 - \vec{r}^2 - u^2$, or simply $\vec{r}^2 + u^2 = t^2$.

\vspace{4mm}
\begin{remark} 
	The manifold ($\pazocal{H}^4$, $\eta_{1,4}$) is spatially homogeneous. The homogeneity of $\pazocal{H}^4$ is verifiable because it can be foliated by spatial hypersurfaces ${\Sigma_t}$ such that, for $\forall p, q  \in {\Sigma_t}$ and $\forall t$, there exists a transformation (diffeomorphism) that moves from point $p$ to point $q$ and leaves the metric invariant. In other words, a spheroidal submanifold ${S_t^3}$ can be defined for each time $t$ as the intersection between the hypercone $\pazocal{H}^4$ and isochronous hyperplane at this time $t$:
	\begin{eqnarray} \label{eq:S3}
	{S_t^3} := \left\{ (\vec{r},u) \in \mathbb{R}^{4} : { \vec{r}}^2 + u^2 = t^2 \right\} \subset \pazocal{H}^4
	\end{eqnarray}
\end{remark}  
\vspace{4mm}

Finally, to obtain the metric tensor $g$ and the corresponding intrinsic viewpoint of the hypercone, moving charts are required to represent comoving observers. 

\subsection{Moving reference frames}
\label{sec:Moving.charts}

An observer living in $\pazocal{H}^4$ will measure the local (differential) distances as in its tangent Minkowskian spacetime $\mathbb{R}^{1,3}_{\eta} := (\mathbb{R}^{4}, \eta_{1,3})$. To check this, one takes an infinitesimal difference between three paths $X,X_0,X_0'\,: \mathbb{R}\to \pazocal{H}^4$ and intersects on the hypersurface $S^3_t \subset \pazocal{H}^4$ in $X(t)$, $X_0(t)$ and $X_0'(t)$, respectively. For example, $X_0$ represents a comoving observer with coordinates $X_0(t) = (t,\vec{0}, t)$  and $X(t) = (t,\vec{r}, u)$ is the image of any arbitrary path on the hypercone $\pazocal{H}^4$, while $X_0'(t)=(t_0, \vec{0}, t)$ is the image of the observer \textbf{reference path} at time $t_0 \in \mathbb{R}_{>0}$.  Notice that the observer and its reference path image are not the same, $X_0 \neq X_0'$, but they coincide for each measurement at the time $t=t_0$, if it is set as a particular constant value. Therefore, the simple differential distances are\begin{eqnarray} \label{eq:sint}
    ds^2 &:=& d(X(t)-X_0(t))^2 = (0,d\vec{r},d(u-t))^2  \\ \label{eq:cont}
     ds'^2 &:=& d(X(t)-X_0'(t))^2 = (dt,d\vec{r},d(u-t))^2\,,\;
\end{eqnarray}that only differ in the temporal coordinate. The last term contributes to the following: \begin{eqnarray} \nonumber
d(u-t)  =d\left( t \cos\gamma-t\right) = - t\sin\gamma d\gamma + (\cos\gamma-1)dt \end{eqnarray} where $\gamma = \arcsin\frac{r}{t}$ , then,\begin{eqnarray} \nonumber
    d(u-t)^2  =  t^2\sin^2\gamma d\gamma^2 + (\cos\gamma-1)^2dt^2 - 2t\sin\gamma  (\cos\gamma-1) dt d\gamma
    \\   \addtocounter{equation}{2} {} \label{eq:u}  
    {} 
\end{eqnarray}where $\sin\gamma = r'/t_0 = r/t$ and $\cos\gamma = \sqrt{1-r'^2/t_0^2}$, with comoving coordinate $r'$ of $r$ at the measuring time $t_0$.  Notice that $d(\sin\gamma) = dr'/t_0 = dr/t-rdt/t^2$ and, therefore, $dr = (t/t_0)dr' + r'/t_0 dt$ and $d{\vec{r}} = d(r \vec {e_r} )= d{r}\vec{e_r} + rd\Sigma \vec{e_{\Sigma}}$. Thus, \begin{eqnarray}
\label{eq:r}  \nonumber 
   d\vec{r}^2 = \left(\frac{t}{t_0}\right)^2(dr'^2+r'^2d\Sigma^2) + \frac{r'^2}{t_0^2} dt^2  + 2\frac{t}{t_0}\frac{r'}{t_0} dr'dt
\end{eqnarray}Now, adding both the spatial term (Eq. \ref{eq:r}) and the u-dimension contribution (Eq. \ref{eq:u}) to Eqs. \ref{eq:sint} and \ref{eq:cont}, they display\begin{eqnarray}
    ds^2 = \left(2\cos\gamma-\mathbf {2}\right)dt^2-\frac{t^2}{t_0^2}\left(\frac{dr'^2}{1-\frac{r'^2}{t_0^2}} + r'^2d\Sigma^2 \right)   \addtocounter{equation}{1} {} \label{eq:gbad} 
    - \; 2\frac{t}{t_0}\frac{r'}{t_0}\frac{dr'dt}{\cos\gamma} \hspace{6mm}
    \\ 
     ds'^2 = \left(2\cos\gamma-\mathbf{1}\right)dt^2-\frac{t^2}{t_0^2}\left(\frac{dr'^2}{1-\frac{r'^2}{t_0^2}} + r'^2d\Sigma^2 \right)
     \addtocounter{equation}{0} {}  \label{eq:ghyp}
     -  \; 2\frac{t}{t_0}\frac{r'}{t_0}\frac{dr'dt}{\cos\gamma}  \hspace{6mm}
\end{eqnarray}where boldface numbers highlight the difference in the temporal contribution. Finally, one can check the local limit for $r' << t_0$ is only Minkowskian for the second equation,\begin{eqnarray} 
    ds^2 \big|_{r' << t_0}& =& -\frac{t^2}{t_0^2}\left(dr'^2 + r'^2d\Sigma^2 \right) 
    \\ 
     ds'^2 \big|_{r' << t_0} &= &dt^2 -\frac{t^2}{t_0^2}\left(dr'^2 + r'^2d\Sigma^2 \right)\,. 
\end{eqnarray}Notice that the metric is only flat in space since Ricci scalar $R = -6\dot{a}^2/a = -6/t^2$ is not zero due to the contribution of the expansion scale factor $a = t/t_0$.

\subsection{Minimal dynamical embedding proposition}

\begin{proposition}[Minimal embedding] \label{prop:hypmetric}
    The hyperconical metric (Eq. \ref{eq:ghyp}) is the minimal dynamical embedding of the hypercone $\pazocal{H}^4 \subset \mathbb{R}^{1,4}$.
    
     \textbf{Proof}. To obtain a four-dimensional metric that inherits the dynamics of $\pazocal{H}^4$, moving charts defined with respect to $X_0'(t)=(t_0, \vec{0}, t)$ adequately represent observers to measure at any time $t_0$. To probe that it is a unique solution, we suppose that another reference is $Y_0'(t) = (\tau(t), \vec \rho(t), \nu(t))$. If $Y_0'(t_0) \in \pazocal{H}^4$, it means that $\tau(t_0)^2 = \vec \rho(t_0)^2 + \nu(t_0)^2$ for each $t=t_0$. Moreover, there exists a coordinate change $Y_0' \to Y_0''$ in which $\vec \rho \to 0$ by simple displacement $\Delta Y = (0,\vec\rho,0)$ and, therefore, the hypercone equation $\tau_0^2 := \tau(t_0)^2 = \nu(t_0)^2$ leads to\begin{itemize}
        \item $Y_0''(\tau(t)) = (\tau_0,\vec 0, \tau)$ or
        \item $Y_0''(\tau(t)) = (\tau,\vec 0, \tau_0)$ or 
        \item $Y_0''(\tau(t)) = (\tau,\vec 0, \tau)$
    \end{itemize} but the last two cases are rejected because they do not reproduce the local Minkowskian distance. Therefore, only the first solution $Y_0''(\tau) = (\tau_0,\vec 0, \tau)$, with ($t \to \tau$)-transformation, is valid; which corresponds to the reference of moving charts for the hyperconical metric of Eq. \ref{eq:ghyp}. \ensuremath{\blacksquare}  
\end{proposition}

\smallskip
  \smallskip

\section{Inhomogeneity and acceleration}
\label{sec:accel}

\subsection{What is it?}

This Appendix summarizes the proof that radial inhomogeneity of the hyperconical metric behaves in a similar way that a fictitious acceleration. That is, it produces the same effect on the redshift measured by an observer, as is detailed in \citep{Monjo2018, MCS2023}. In other words, there exist coordinate changes that transform the hyperconical model in an apparent accelerated flat FLRW universe.

\subsection{Radial inhomogeneity}

Terms of lapse and shift of the hyperconical metric (Eq. \ref{eq:ghyp}) can be assimilated as a radial inhomogeneity according to new coordinates. Applying a coordinate change in time $t \to t'$ to the metric, such as $t' := t\sqrt{2\cos\alpha-1}$, it is equivalent to selecting $g_{00}' = 1$, $g_{0r}' = 0$ and:
\begin{eqnarray} \label{eq:grr}   
 g_{r'r'}' &=& g_{r'r'} - \frac{g_{0r}^2}{g_{00}} = -{a(t',r')}^2 \frac{2-\cos\gamma(r')}{2\cos^2\gamma(r')} \\ \label{eq:gff}   
 g_{\phi \phi}'  &= &-{a(t',r')}^2 {r'}^2 sin^2 \theta \\ \label{eq:ghh} 
 g_{\theta \theta}'& = &-{a(t',r')}^2 {r'}^2            
\end{eqnarray} where it is assumed the positive curvature $k=1$ as it was found in \citep{MCS2020}, and $a(t',r') = t'/(t_{\hat{o}}\sqrt{2\cos\gamma-1})$ is the same scale factor but expressed in the new coordinates. From the definition of luminosity distance \citep{Monjo2017}, the comoving distance $r'$ is derived as a function of redshift $z$, \begin{eqnarray} \label{eq:comoving.distance2}
 \mathfrak{sn}_{1}^{-1} \left( \frac{r'}{t_0} \right) :=   \int_{0}^{r'} \frac{\sqrt{2-\cos\gamma(r')}}{\cos\gamma(r')} \frac{d r'}{t_0}  & = & \int_0^z \frac{d z'}{1 + z} \;\;\;\;\;\;\;\;\;\;
 \\  \nonumber 
 \int_{0}^{\gamma(r')} \sqrt{2-\cos\gamma} \,d \gamma   & = &  \ln(1 + z) 
\\ \addtocounter{equation}{1}  \label{eq:equivalz}
\gamma(r') + O[\gamma^3]  & = & z + O[z^2] \;\;\;\;\;\;\;\;\;\;
 \end{eqnarray} Isolating the comoving distance $r'$ and applying a locally conformal projection map $f_{\gamma_0}$ (e.g. \ref{eq:projectionmap}) to change this coordinate to the intrinsic distance $\hat{r}'$, it is the same that:
\begin{eqnarray}\label{eq:comoving.distance3}
r' &= &{t_0} \mathfrak{sn}_{1}( \ln(1 + z) )
\\ \label{eq:comoving.distance4}
\hat{r}' & = & f_{\gamma_0}(r')
\end{eqnarray}The hypothesis is that intrinsic (projected) distances of the linear expanding and closed hyperconical model behaves like the accelerated and flat $\Lambda$CDM model. Taking the Maclaurin series on redshift for the Hubble parameter of both models ($H_{\Lambda CDM}$ and $\hat{H}_{hyp}^{intr}$ ), the first order is \begin{eqnarray} \label{eq:hubble_lcdm}  \nonumber
H_{\Lambda CDM\ }(z) &=& 
 \sqrt{\Omega_{r} +\Omega_{m} + \Omega_{\Lambda}}
+ \frac{4\Omega_{r} + 3\Omega_m}{\sqrt{\Omega_{r} + \Omega_{m} + \Omega_{\Lambda}}} \frac{z}{2} +
\\ \nonumber
&& \;  + O(z^2) 
\\  \addtocounter{equation}{2} {}    \label{eq:hubble_hyp} 
  \hat{H}_{hyp\ }^{intr}(\hat{r}'(z)) &=& \hspace{1mm} 1 \hspace{1mm}  + \hspace{1mm}  \frac{\gamma_0 - 2\alpha}{\gamma_0} z \hspace{1mm} + \hspace{1mm} O(z^2) 
\end{eqnarray}where $\alpha = 0.5$ is the unique distorting parameter compatible under symmetric transformation in dynamical systems \citep{MCS2023}. 

\begin{proposition}[\textbf{Inhomogeneity--acceleration equivalence}] Given the redshift measured by comoving observers in the linearly expanding hyperconical metric, there exists a unique local equivalence (stereographic projection) between radial inhomogeneity of that metric and an acceleration of the spatially flat FLRW metric. \label{prop:inhomo-accel}

\textbf{Proof}. Let $\lambda \subset \mathbb{R}$ be a scale factor that transforms time $t \mapsto \hat{t} := t\lambda$ and distorts the comoving length $\hat r'$ as $r' \mapsto \hat r' := r'\lambda^\alpha$, with $r' = | \vec{r}'|$ for each point of the submanifold $S_t^3 = ((t, \vec{r}', u) \in \mathbb{R}^5\;:\; | \vec{r}'|^2 + u^2 = t^2)$. Moreover, let $\{F_Q : \mathbb{R} \to \mathbb{R}^5\}_{Q \in S_{t}^3}$ be a family of map projections with $F_Q(\lambda) \in \mathbb{R}^5$ parameterized, such as $F_Q(1) = Q(t) = (t, \vec{r}, u)$ and $F_Q(0) = \hat Q(0)  := (0, \vec{0}, u_0)$ with $u_0 := - t_0$. A transformation, given by $\vec{r}' = r'\vec{e}_{r} \; = \;  t_0 \sin  \gamma\, \vec{e}_{r} \; \mapsto \; \hat r'\vec{e}_{r} := t_0 \sin \hat \gamma\, \vec{e}_{r}$, is performed for the angles $\gamma \mapsto \hat \gamma$, preserving the direction $\vec{e}_{r}$. Therefore, that family of maps is:\begin{eqnarray}
 \nonumber
F_{Q}
\begin{cases}
\hat t =  t \lambda \\ \nonumber 
\hat  r' =  r'\lambda^{\alpha}\\
\hat u = u_0 + (u - u_0) \lambda 
\end{cases} 
\end{eqnarray}where $\alpha = 0.5$ \citep{MCS2023}. When the points are projected on the hyperplane $\hat u = t_0$, the solution for the distorted stereographic projection is given by some $\lambda = \lambda_s(t,\gamma)$, \begin{eqnarray} \nonumber 
t_0 = - t_0 + \left( t\cos\gamma  +  t_0 \right)\lambda_s \;\; \Rightarrow\;\;
\lambda_s(t,\gamma) = \frac{2}{(1+\frac{t}{t_0}\cos\gamma)} \\ \addtocounter{equation}{2} {}  \label{eq:lambda}
{}
\end{eqnarray} Taking into account the inhomogeneous scale factor  related to redshift $z$ as a function of the comoving distance (Eq. \ref{eq:equivalz}), \begin{eqnarray} \nonumber
 \frac{t}{t_0} &= & \frac{1}{1+z} = 1 - \frac{r'}{t_0} +  O\left({\frac{r'^2}{t_0^2}}\right) = 
 \\  \addtocounter{equation}{1} {}   \label{eq:localtime}
 &=& 1 - \sin{\gamma} +  O\left(\sin^2{\gamma}\right)  = 1 - \gamma +  O\left(\gamma^2\right) \,,\;\;\;
 \end{eqnarray}the projection parameter can be found:
 \begin{eqnarray} \label{eq:lambda_local0}
 t_0 &= & - t_0 + \left( t\cos\gamma  +  t_0 \right)\lambda_s \;\; \Rightarrow\;\; 
 \lambda_s(t(\gamma), \gamma) \approx \frac{2}{(1+\left(1 - \gamma \right)\cos\gamma)}
 \end{eqnarray}At local scales ($\gamma << 1$), it is useful to approximate Eq. \eqref{eq:lambda_local0} as\begin{eqnarray} \label{eq:lambda_local}
      \lambda_S(t(\gamma), \gamma) \approx \frac{1}{1 - \frac{\gamma}{2}}
\end{eqnarray}Now, taking the Eq.~\ref{eq:hubble_hyp} for the local value of $\gamma_0 = 2$, \begin{eqnarray}
    \hat{H}_{hyp\ }^{intr}(\hat{r}'(z))  \hspace{1mm} \approx \hspace{1mm}  1 \hspace{1mm}  + \hspace{1mm} \frac{1}{2} z + O(z^2) 
\end{eqnarray}and comparing to the Hubble parameter for the spatially flat FLRW metric,\begin{eqnarray}
    \begin{cases}
         1 &= \sqrt{\Omega_{r} +\Omega_{m} + \Omega_{\Lambda}} \\
         \frac{1}{2} z & =  \frac{4\Omega_{r} + 3\Omega_m}{\sqrt{\Omega_{r} + \Omega_{m} + \Omega_{\Lambda}}} \frac{z}{2}
    \end{cases}
\end{eqnarray} 

    \smallskip
   
Therefore, $\{\Omega_m = \frac{1}{3}-\frac{4}{3}\Omega_r,\;\Omega_\Lambda = \frac{2}{3} + \frac{1}{3}\Omega_r\}$ is the unique local solution as a function of the radiation density ($\Omega_r$). Simplifying with $\Omega_r \approx 0$, we obtain $\Omega_m \approx \frac{1}{3}$ and $\Omega_\Lambda \approx \frac{2}{3}$ under the first-order approach. \ensuremath{\blacksquare}   
Second- \citep{Monjo2018} and third- \citep{MCS2023} order approaches are not unique, but display similar results.    
\end{proposition}

\begin{theorem}[Fictitious acceleration]
    \label{prop:ficticious} If $M \in \mathbb{R}_\eta^{1,4}$ is a minimal singular (or maximal regular) four-dimensional hypersurface with finite parameter time, every comoving on-shell observer will experience a local fictitious acceleration.

    \textbf{Proof}. By the Prop.~\ref{prop:minsingular}, maximal regularity of $M$ implies that it is closed and, by homogeneity, $M$ is an expanding hypersphere. Moreover, from Prop.~\ref{prop:hypmetric}, embedding of $M$ is unequivocally described by the linearly expanding hyperconical metric (except under coordinate changes), which presents a radial inhomogeneity. Finally, according to Prop.~\ref{prop:inhomo-accel}, local projection of the inhomogeneity leads to a fictitious acceleration measured by every comoving observer. \ensuremath{\blacksquare}  
                \end{theorem}

\smallskip
  \smallskip
  \smallskip
  
\section{Data used to fit expansion rate}
\label{sec:dataCC}

The temporal evolution of the expansion, fitted in Fig.~\ref{fig:Fig3}, uses 51 pairs of Hubble parameter ($H$) and redshift ($z$) collected from multi-source datasets (Table \ref{tab:B1}): cosmic chronometers (CC), distance ladder, and radial BAO size from several approaches: galaxy distribution (BAO-Gal), quasars (BAO-QSO, Lymann-$\alpha$ forest (BAO-Ly$a$).


\begin{table}[]
    \centering
     \caption{\label{tab:B1}Hubble parameter ($H$) and statistical error ($\sigma$) as a function of redshift ($z$), estimated by using cosmic chronometers (CC: 34), type-Ia SNe distance ladder (Ladder: 3) and radial BAO size from several approaches: Galaxy distribution (BAO-Gal: 7), redshift space distortions from eBOSS quasars (BAO-QSO: 4) and Lymann-$\alpha$ forest (BAO-Ly$a$: 3).}
    \label{tab:my_label}
    \begin{tabular}{ccccc}
   \hline
    \hline
    z & $H(z) / km s^{-1} Mpc^{-1}$ & $\pm 1\sigma(H(z))$ & Method & Reference  \\
		\hline
$<$0.011 & 73.3 & 1.04 & Ladder & \citet{Riess2022} \\
\quad 0.023-0.15 \quad\quad\quad & 73.32 & 0.99 & Ladder & \citet{Riess2022} \\	
0.0708 & 69.0 & 19.68 & CC & \citet{Zhang2014}\\
0.09 & 69.0 & 12.0 & CC & \citet{Jimenez2003}  \\
0.12 & 68.6 & 26.2 & CC & \citet{Zhang2014}  \\
0.17 & 83.0 & 8.0 & CC & \citet{Simon2005} \\
0.179 & 75.0 & 4.0 & CC & \citet{Moresco2012}  \\
0.199 & 75.0 & 5.0 & CC & \citet{Moresco2012} \\
0.2 & 72.9 & 29.6 & CC & \citet{Zhang2014}  \\
0.240 & 79.69 & 2.65 & BAO-Gal & \citet{Gaztañaga2009}\\
0.27 & 77.0 & 14.0 & CC &\citet{Simon2005}  \\
0.28 & 88.8 & 36.6 & CC & \citet{Zhang2014}   \\
0.35 & 84.4 & 7.0 & BAO-Gal & \citet{Xu2013} \\
0.352 & 83.0 & 14.0 & CC & \citet{Moresco2012} \\
0.3802 & 83.0 & 13.5 & CC &  \citet{Moresco2016}  \\
0.4 & 95.0 & 17.0 & CC & \citet{Simon2005}  \\
0.4004 & 77.0 & 10.2 & CC  &  \citet{Moresco2016} \\
0.4247 & 87.1 & 11.2 & CC &  \citet{Moresco2016}\\
0.43 & 86.45 & 3.68 & BAO-Gal & \citet{Gaztañaga2009}\\
0.44 & 82.6 & 7.8 & BAO-Gal & \citet{Blake2012}  \\
0.4497 & 92.8 & 12.9 & CC & \citet{Moresco2016}  \\
0.47 & 89 & 50 & CC & \citet{Ratsimbazafy2017}  \\
0.4783 & 80.9 & 9.0 & CC  &  \citet{Moresco2016}  \\
0.48 & 97.0 & 62.0 & CC & \citet{Stern2010}  \\
0.57 & 92.4 & 4.5 & BAO-Gal & \citet{Samushia2013} \\
0.593 & 104.0 & 13.0 & CC & \citet{Moresco2012} \\
0.6 & 87.9 & 6.1 & BAO-Gal & \citet{Blake2012}  \\
0.68 & 92.0 & 8.0 & CC &  \citet{Moresco2012}  \\
0.73 & 97.3 & 7.0 & BAO-Gal & \citet{Blake2012}  \\
0.75 & 98.8 & 33.6 & CC & \citet{Borghi2022}   \\
0.8 & 113.1 & 15.1 & CC & \citet{Jiao2023} \\
0.781 & 105.0 & 12.0 & CC & \citet{Moresco2012}  \\
0.875 & 125.0 & 17.0 & CC & \citet{Moresco2012} \\
0.88 & 90.0 & 40.0 & CC & \citet{Stern2010} \\
0.9 & 117.0 & 23.0 & CC & \citet{Simon2005}  \\
0.978 & 113.72 & 14.63 & BAO-QSO & \citet{Zhao2018}\\
1.037 & 154.0 & 20.0 & CC & \citet{Moresco2012} \\
1.23 & 131.44 & 12.42 & BAO-QSO & \citet{Zhao2018} \\
1.26 & 135 & 65 & CC & \citet{Tomasetti2023}  \\
1.3 & 168.0 & 17.0 & CC & \citet{Simon2005} \\
1.363 & 160.0 & 33.6 & CC &  \citet{Moresco2015} \\
1.43 & 177.0 & 18.0 & CC & \citet{Simon2005} \\
1.5 & 197 & 50 & Ladder & \citet{Riess2018} \\
1.526 & 148.11 & 12.75 & BAO-QSO & \citet{Zhao2018}\\
1.53 & 140.0 & 14.0 & CC & \citet{Simon2005} \\
1.75 & 202.0 & 40.0 & CC & \citet{Simon2005} \\
1.944 & 172.63 & 14.79 & BAO-QSO & \citet{Zhao2018}\\
1.965 & 186.5 & 50.4 & CC & \citet{Moresco2015} \\
2.33 & 224 & 8 & BAO-Ly$\alpha$ & \citet{Bourboux2017} \\
2.34 & 222.0 & 7.0 & BAO-Ly$\alpha$ & \citet{Delubac2015}  \\
2.36 & 226.0 & 8.0 & BAO-Ly$\alpha$ & \citet{FontRibera2014} \\
    \hline
    \end{tabular}
\end{table}

    \newpage


\bibliography{main}{}
\bibliographystyle{aasjournal}



\end{document}